\documentclass[
aps,
prl,
reprint,
superscriptaddress,
twocolumn,
longbibliography
]{revtex4-1}

\usepackage{graphicx}
\usepackage{dcolumn}
\usepackage{bm}
\usepackage{amsmath,amssymb,amsthm}

\newtheorem{thm}{Theorem}
\newtheorem{lem}[thm]{Lemma}
\newtheorem{cor}[thm]{Corollary}

\begin{document}


\title{\bf\huge Computational Characteristics of Random Field Ising Model with Long-Range Interaction}


\author{Fangxuan Liu}
\email{liufx19@mails.tsinghua.edu.cn}
\affiliation{Center for Quantum Information, Institute for Interdisciplinary Information Sciences, Tsinghua University, Beijing 100084, PR China}

\author{L.-M. Duan}
\email{lmduan@tsinghua.edu.cn}
\affiliation{Center for Quantum Information, Institute for Interdisciplinary Information Sciences, Tsinghua University, Beijing 100084, PR China}

\date{\today}

\begin{abstract}
Ising model is a widely studied class of models in quantum computation. In this paper we investigate the computational characteristics of the random field Ising model (RFIM) with long-range interactions that decays as an inverse polynomial of distance, which can be achieved in current ion trap system. We prove that for an RFIM with long-range interaction embedded on a 2-dimensional plane, solving its ground state is $\mathsf{NP}$-complete for all diminishing exponent, and prove that the 1-dimensional RFIM with long-range interaction can be efficiently approximated when the interaction decays fast enough.
\end{abstract}

\maketitle


\section{Introduction}

Quantum computer science are in the noisy intermediate-scale quantum (NISQ) era~\cite{NISQ}, when it is still not possible to build a large-scale quantum computer with capable error correction, but it is already possible to build a quantum circuit with intermediate scale (tens of qubits) and depth (tens of steps) with relatively high fidelity~\cite{google_supercomputer,utsc_superconductor,threshold_1,threshold_2}.

While the full power of quantum computation, such as solving computationally hard problems~\cite{quantum_algorithms}, applications on optimization~\cite{variational_quantum_algorithm}, quantum machine learning~\cite{quantum_machine_learning}, quantum chemistry~\cite{quantum_chemistry} and many other fields can only be fulfilled beyond the NISQ era, analogue quantum computing~\cite{analogue_quantum_computation_1,analogue_quantum_computation_2,analogue_quantum_computation_3} thrives in this circumstance. Instead of using sequence of gates dictated by quantum algorithms to manipulate qubits, it simulates the target system by manipulating the Hamiltonian of a well controlled quantum system to solve computational problems. Although important progress has been achieved in this field, currently the Hamiltonian that can be simulated is often restricted by the specific physical systems and the available experimental techniques.

Ising model~\cite{ising_history} is among the systems that are most frequently used in analogue quantum computing because of its simplicity and capability. The computational complexity problem associated with Ising models is a widely discussed topic, as its result will affect the utilization of different families of Ising models in analogue quantum computing field. The general Ising model is well-known to be NP-complete~\cite{related_ising_complexity_2}. Indeed Karp's 21 NP-complete problems~\cite{independent_set_npc} can reduce to the ground state of Ising models~\cite{ising_formulation_npc}, and it is also connected to the complexity of general high-spin systems~\cite{spin_system}. For more specific families of Ising models, there are also many known results on their complexity. Ising model without magnetic field on a planar graph is proven to be in $\mathsf{P}$ \cite{related_ising_complexity_1}, while Ising model without field on non-planar graph is $\mathsf{NP}$-complete \cite{related_ising_complexity_2}. Ising model with random fields are in $\mathsf{P}$ when the coupling is ferromagnetic, and is in $\mathsf{NP}$-complete when the coupling is anti-ferromagnetic in 2 dimensions or more ~\cite{related_ising_complexity_3}. Ref.~\cite{spin_glass_3d} gives a further examination about the lower bound of the complexity of 3D Ising model with magnetic field. These previous research works either consider the most general Ising model, or very specific models in terms of the interaction and the field terms, which are challenging for current experiment to construct.

The Ising models that naturally appears in the ion trap are mostly random field Ising models with long-range interactions: the interaction between spins can usually be approximated by the form $r^{-\alpha}$~\cite{poly_diminish_1,poly_diminish_2}, with $r$ being the distance between the two spins and $\alpha$ being a constant in the range of $0-3$ depending on the frequency detuning of the operation laser beams. Hence, it is crucial to study the computational characteristics of this family of Ising models.

In this paper, we prove several computational characteristics of long-range interaction random field Ising models (RFIM)~\cite{random_field_ising_model}. In section~\ref{sec:notation}, we give the background of our proof and define the notations. In section~\ref{sec:outline} we gives the outline of the proof. In section~\ref{sec:NPC}, we prove that the 2D grid anti-ferromagnetic long-range interaction RFIM is in $\mathsf{NP}$-complete. In section~\ref{sec:APX}, we consider Ising model as an optimization problem, prove that the 1D long-range interaction RFIM have a Polynomial-Time Approximation Scheme (in complexity class $\mathsf{PTAS}$), and hypothesize that the 2D long-range interaction RFIM is not likely in $\mathsf{PTAS}$.

\section{Background and Notations}\label{sec:notation}

A Random Field Ising Model (RFIM) is an undirected graph $G=(V,E)$ together with a Hamiltonian $H$ of the form:

$$
H=-\sum_{\{i, j\}\in E} J_{ij} \sigma_i \sigma_j - \sum_{k\in V} h_k \sigma_k
$$

A vertex of the graph is called a \emph{spin}. The \emph{state} $\sigma$, assigns a value $\pm 1$ for each spin. The \emph{interaction}, assigns a real value $J_{ij}$ for each edge $\{i, j\}\in E$. The \emph{field}, assigns a real value $h_{i}$ for each spin $i$.

Given the interaction and field, the Hamiltonian gives the energy of a state: $E(\sigma) = H(\sigma;J,h)$. A state is called \emph{ground state} if there exists no state that have lower energy.

Now we consider Ising model that is embedded onto Euclidean space. Each spin $v$ is assigned to a position in the Euclidean space $\vec{r}_i$, and the interaction $J_{ij}$ is dictated by the position of two spins: $J_{ij} = f(\vec{r}_i, \vec{r}_j)$.

Here we consider a specific family of $f$: $f(\vec{r}_i, \vec{r}_j)=C||\vec{r}_i-\vec{r}_j||^{-\alpha}, \alpha>0$, the interaction of two spins are determined by their distance, and this system is ferromagnetic (anti-ferromagnetic) if $C$ is positive (negative). This family of interaction naturally occurs in ion traps, and is the most common family of interaction that is able to be prepared in the current laboratory. We call this family of interaction \emph{long-range interaction}, and we say that this Ising model have $\alpha$-long-range interaction.

\section{Outline of the Proof}\label{sec:outline}

The main idea of the proof of NP-completeness of the 2D grid anti-ferromagnetic long-range interaction RFIM is to construct several systems that have different exponent for the long-range interaction, and construct maps between the state of two systems. These systems and mappings have special properties so that the mapping only introduce minor relative perturbation to the energy states. The composition of the mappings connects an realistic system (an Ising model with relatively long range interaction) and an ideal system (an Ising model with perfect near neighbour interaction), and if there exist an oracle that gives the ground state of the ideal system, we can utilize the oracle and the mapping, which can be executed in polynomial time, to solve an NP-complete problem.

We also examine the computation complexity of the one dimensional system by giving a proper definition of $\epsilon$-approximation of the solution and give a dynamic-programming styled algorithm for efficient approximation. We further expand the set of Ising model that can be efficiently approximated to Ising models that satisfies a specific graph structure, and give an conjecture of the non-approximatable condition of the Ising model.

\section{NP-completeness of 2D grid Anti-ferromagnetic Long-Range Interaction RFIM}\label{sec:NPC}

Here we prove the theorem that solving the ground state of 2D grid Anti-ferromagnetic Long-Range Interaction RFIM is $\mathsf{NP}$-complete.

\subsection{Low energy state preserving maps}

Here we consider two Ising models and a map between the states of two models.

For an Ising model, we define a set $G$ of states a \emph{low-energy set} if: there exist no state $\sigma'\not\in G$ so that $\exists \sigma\in \sigma', E(\sigma')<E(\sigma)$. In particular, the set that contains all ground state of an Ising model is a low-energy set.

The \emph{energy gap} $\Delta_G$ of a low-energy set characterizes how far this set is from the other states: $\Delta_G=\min_{\sigma\not\in G} E(\sigma) - \max_{\sigma\in G} E(\sigma)$.

Now we consider two Ising models with states $\Sigma = \{\sigma\}$ and $\Sigma' = \{\sigma'\}$. Let $\tilde{\Sigma}$ be a subset of $\Sigma$ and $\tilde{\Sigma'}$ be a subset of $\Sigma'$. Consider a bijective map $g:\tilde{\Sigma'} \rightarrow \tilde{\Sigma}$, we call it a $(a,\delta)$ map if:

$$
\forall \sigma'\in G, |E(\sigma') - aE(g(\sigma'))| \le a\delta
$$

This map preserves the low-energy set under the condition that:

\begin{lem}\label{thm:gap}
For a low-energy set $G\subseteq \tilde{\Sigma}$, if:
\begin{enumerate}
    \item $\delta<\frac{1}{2}\Delta_G$.
    \item $\Tilde{\Sigma'}$ is a low-energy set.
\end{enumerate}
Then, $g^{-1}(G)$ is a low-energy set with energy gap $\Delta_{g^{-1}(G)} = a(\Delta_G-2\delta)$
\end{lem}

Besides, we consider the composition of two maps:

\begin{lem}\label{thm:composite}
Consider three systems with states $\Sigma, \Sigma', \Sigma''$, let $\tilde{\Sigma},\tilde{\Sigma'},\tilde{\Sigma''}$ be the subset of $\Sigma, \Sigma', \Sigma''$ respectively. Consider two maps $g:\tilde{\Sigma'}\rightarrow  \tilde{\Sigma}$ and $g':\tilde{\Sigma''} \rightarrow \tilde{\Sigma'}$. If $g$ is a $(a,\delta)$-map and $g'$ is a $(a',\delta')$-map, then the composition $g\circ g'$ is a $(a'a,\frac{\delta'}{a}+\delta)$-map.
\end{lem}

\subsection{Arbitrary placement of spins}

Here we consider how it will affect the energy states of an Ising model if we slightly change the position of its spins. The result shows that although we are limited to put spins on the grid point, we can construct a system with arbitrarily small error where we can place the spins arbitrarily.

\begin{lem}\label{thm:mapgrid}
Consider an Ising model with $\alpha$-long-range interaction (original system) on 2D space with $n$ spins, with each spin $i$ located on $\vec{r}_i =(x_i,y_i)$. The minimal distance between two spins are $d$. Then, consider another Ising model with $\alpha$-long-range interaction (new system) on 2D space, with:

For each spin $i$, we put a spin $i'=(\lfloor\frac{x_i}{\epsilon}\rfloor,\lfloor\frac{y_i}{\epsilon}\rfloor)$, and apply the magnetic field $h_{i'} = h_i\epsilon^\alpha$.

Then we construct a map $g:\Sigma'\rightarrow \Sigma$: A state in the new system $\sigma'$ is send to $\sigma$ in the original system, with $\sigma_{i}=\sigma'_{i'}$ for all spin $i$ in the original system.

With $\epsilon<\frac{d^{\alpha+1}\delta}{2(\alpha+1)n(n+1)}$, $g$ is a $(\epsilon^\alpha, \delta)$-map.
\end{lem}

This theorem implies that we can simulate an Ising model with spins placed on arbitrary system with an Ising model with spins placed on square grid with arbitrary precision with the same spin count.

\subsection{Logical spins}

Consider an anti-ferromagnetic Ising model with an $\alpha$-long-range interaction that has 8 spins. Four of the spin are placed on $(\pm 1, \pm 1)$, and four of the spins are placed on $(0,\pm1)$ and $(\pm 1,0)$. With no magnetic field applied to all spins, this Ising model have two ground states.

The ground state of this system is 2-degenerate: Either the corner spins are $+1$, the edge spins are $-1$ or the opposite. We use $\sigma_+$ and $\sigma_-$ to denote these two states. The energy gap between these two ground states and the other states are determined by $\alpha$: $\Delta(\alpha) = 4-4\cdot 2^{-\frac{\alpha}{2}} - 6\cdot 4^{-\frac{\alpha}{2}} + 8\cdot 5^{-\frac{\alpha}{2}} + 2\cdot 8^{-\frac{\alpha}{2}}$. The energy gap is a constant given the $\alpha$ being constant.

We call $\sigma_\pm$ \emph{valid states}, and if we only consider the valid state of the system, we can use a single variable $\tilde{\sigma}$ to denote the two states: $\tilde{\sigma}=1$ indicates the system is in $\sigma_+$ state, and $\tilde{\sigma}=-1$ indicates the system is in $\sigma_-$ state.

Also, if we apply magnetic field $+\frac{h}{8}$ on the corner spins and $-\frac{h}{8}$ on the edge spins, then $E(\sigma_-)-E(\sigma_+)=2h$, we call this field scheme \emph{logical field} $h$. Then, if we only consider the valid states of this model, this model is equivalent to the model that contains the singleton spin and with magnetic field $h$ applied to it, hence we call this model \emph{logical spin}. For a model that have multiple logical spins, a state is called valid if all the logical spins are in valid states.

The following theorem gives the condition that in which situation the set of valid state is a low-energy set given the other spins fixed:

\begin{lem} \label{thm:valid}
Consider an Ising model with a logical spin with logical field $+h$ on the origin and $n$ spins and the distance from the origin to each spin greater than $R = (\frac{1}{8n}(\frac{\Delta}{2}-|h|))^{-\frac{1}{\alpha}} + \sqrt{2}$. Then if all these spins are fixed, $\sigma_\pm$ are the two lowest-energy state among all states of this fields.
\end{lem}

So, if several logical spins are placed in the space and they are not too close to each other, we can expect that if the state of a logical spin is not one of $\sigma_\pm$, then it is impossible to be the ground state.

Next, we consider the property of the interaction between two logical spins that satisfies the above condition.

\begin{lem}\label{thm:interaction}
Consider an Ising model with two logical spins that is placed satisfying Lemma~\ref{thm:valid}, the distance between them are $r$. The system can be described as $\Tilde{\sigma} = (\Tilde{\sigma}_1, \Tilde{\sigma}_2)$ ignoring the invalid state. We apply $h_1,h_2$ logical field onto the two logical spins respectively. Up to a constant, the Hamiltonian of the system is:

$$
H(\Tilde{\sigma}) = -h_1\Tilde{\sigma}_1 -h_2\Tilde{\sigma}_2 - I_{12} \Tilde{\sigma}_1\Tilde{\sigma}_2
$$

And $I_{12}$ satisfies:

$$
\left|I_{12} + \alpha^2 (\alpha+2)^2 r^{-(\alpha+4)}\right| \le f(\alpha) r^{-\alpha-5}
$$
\end{lem}

This theorem indicates that if the system has a $\alpha$-long-range interaction, the interaction between two logical spins will be similar to that between two physical spins with a $(\alpha+4)$-long-range interaction. So we can simulate a system with $\alpha+4$-long-range interaction and $n$ spins on a system with $\alpha$-long-range interaction and $8n$ spins:

\begin{lem}\label{thm:maplogical}
Consider an anti-ferromagnetic Ising model with $(\alpha+4)$-long-range interaction and $n$ spins that satisfies $\max_{v\in G} |h_v| \le 1$. Then with parameter $\epsilon$, we can construct an Ising model with $\alpha$-long-range interaction by:

For each spin of original system on $(x,y)$ and magnetic field $h$, we put a logical spin on $(\frac{x}{\epsilon},\frac{y}{\epsilon})$, and apply logical field $h \alpha^2(\alpha+2)^2 \epsilon^{\alpha+4}$ on it.

Then, with $\epsilon$ small enough, Theorem~\ref{thm:valid} is satisfied, so we can only consider the valid states. We can construct a map $g$ from the valid state of the new system to the original system by setting $g(\tilde{\sigma})_i = \tilde{\sigma}_i$.

With $\epsilon < f(n,\epsilon) \delta$, $g$ is a $(\alpha^2(\alpha+2)^2 \epsilon^{\alpha+4}, \delta)$-map.
\end{lem}

\subsection{Polynomial reduction}

In this section, we focus on the reduction from the ground state problem into a known $\mathsf{NP}$-complete problem. We consider this particular problem: find an maximum independent set on a graph with each node on a 2D plane, where an edge connects two nodes if and only if their distance is $1$ and the maximum-degree of each node is 3. Since each node has at most 3 neighbours, we can make the angle of two edges on the same node be greater than $\frac{\pi}{2}$, so if two nodes are not connected, the distance between them is greater than $\sqrt{2}$.

This problem is $\mathsf{NP}$-complete, following directly from the $\mathsf{NP}$-completeness of finding an maximum independent set of a planar graph with max degree of each node being 3.

From \cite{related_ising_complexity_1}, there is:

\begin{lem}\label{thm:nn}
Following \cite{related_ising_complexity_1}, for a planar graph $G=(V,E)$ with max degree 3, we consider an Ising model on this graph with Hamiltonian:

$$
H = \sum_{\{u,v\}\in E} \sigma_u\sigma_v + \sum_{w\in V} \sigma_w
$$

Then this graph has an independent set of cardinality $\ge k$ if there exists a state that $H(\sigma) \le \frac{1}{2}|V|-4k$. Also, from the form of this Hamiltonian, the energy gap of a low-energy set of this system is at least $1$.
\end{lem}

We can find the connection between two Ising models, one built directly on the graph, the other has spins placed on each node, but the interaction between the spins being a long-range interaction with exponent large enough.

\begin{lem}\label{thm:upperbound}
Consider a planar graph $G=(V,E)$ with max-degree 3, each node assigned to a coordinate, with two nodes are connect if and only if their distance is 1, and not connected if and only if their distance is $\ge \sqrt{2}$.

We construct two Ising models:

\begin{enumerate}
    \item An Ising model whose Hamiltonian is:
    $$H = \sum_{\{u,v\}\in E} \sigma_u\sigma_v + \sum_{w\in V} +\sigma_w$$
    \item An anti-ferromagnetic $\alpha$-long-range interaction Ising model with a spin $\tilde{v}$ put on the coordinate of each node and a constant magnetic field $1$ applied on each spin.
\end{enumerate}

Then, we can construct a map $g$ from the second system to the first system by $g(\sigma)_v=\sigma_{\tilde{v}}$. With $\alpha>\frac{2}{\ln 2} \ln (|V|(|V|+1))+2$, this map is a $(1,\frac{1}{4})$-map.
\end{lem}

We can finally prove that finding the ground state of a random field Ising model with anti-ferromagnetic long-range interaction is $\mathsf{NP}$-complete by compositing all the maps constructed above:

\begin{thm}
Given a planar graph $G=(V,E)$ with max-degree 3, each node assigned to a coordinate, with two nodes are connect if and only if their distance is 1, and the distance between nodes are either exactly $1$ or greater than $\sqrt{2}$. Then, for any $\alpha$, there exist an integer $t$ that $\alpha+4t>\frac{2}{\ln 2}\ln(|V|(|V|+1))+2$. We choose the smallest $t$. The size of the problem is characterized by $n$.

\begin{enumerate}
    \item Construct an Ising model according to Lemma~\ref{thm:nn}, denoted by $M_{\text{NN}}$.
    \item Construct the second system, an RFIM with $(\alpha+4t)$-long-range interaction, according to Lemma~\ref{thm:upperbound}, denoted by $M_t$. Call the mapping obtained by this construction $g_{\text{NN}}$, which is a $(1,\frac{1}{4})$-mapping.
    \item For $i$ going from $t$ to $1$, construct an RFIM with $(\alpha+4(i-1))$-long-range interaction $M_{i-1}$ according to Lemma~\ref{thm:maplogical} from $M_{i}$. The map $g_i$ obtained by this construction is a $(F_i, \delta_i)$-map, where $F_i$ is determined by $n,i,\alpha$ and $\delta_i$, and we choose $\delta_i$ so that $\delta_i<\frac{1}{12t\prod_{i<j\le t} F_t}$, this is possible because $\delta_i$ can be chosen arbitrarily small, and the dependency of variables form an acyclic graph.
    \item Construct a RFIM $M_\text{Grid}$ with $\alpha$-long-range interaction and each spin placed on the grid from $M_0$ according to Lemma~\ref{thm:mapgrid}. Choose the parameter that the map $g_\text{Grid}$ obtained by the construction is a $(\epsilon^\alpha, \delta)$-map with $\delta<\frac{\prod_i F_i}{12}$.
\end{enumerate}

Then, the composite map $\tilde{g} = g_\text{NN} \circ g_t\circ \cdots \circ g_1 \circ g_\text{Grid}$ is a $(A, \Delta)$-map with $\Delta < \frac{5}{12}$ from Theorem~\ref{thm:composite}. Because the energy gap of the set of ground states of $M_\text{NN}$ is $1$, its preimage of $\tilde{g}$ on the state of $M_\text{Grid}$ is a low-energy set, so if we find a ground state $\sigma_0$ of $M_\text{Grid}$, $\tilde{g}(\sigma_0)$ is a ground state of $M_\text{NN}$ (Theorem~\ref{thm:gap}).

The number of spins of $M_\text{Grid}$ is $n\times 8^t=O(n(8^{\frac{1}{\ln 2}})^{\ln n}) = O(n^4)$, and the computation of $\tilde{g}$ requires $\text{poly}(n)$ time. So, if we are given a poly-time oracle that gives a ground state of a square-grid anti-ferromagnetic $\alpha$-long-range RFIM, then we can construct a poly-time oracle that gives an maximum independent set of the graph. So, there exist a polynomial time reduction from the ground state of square-grid anti-ferromagnetic $\alpha$-long-range RFIM to an $\mathsf{NP}$-complete problem, indicating that this problem is $\mathsf{NP}$-complete.
\end{thm}

\section{Approximability of Long-Range Interaction RFIM}\label{sec:APX}

Here we view the computational property of the RFIM on a different perspective, as we try to obtain a state whose energy is close to the ground state energy.

This type of problem are usually characterized as optimization problems, and we prove that the approximation of the one dimension RFIM is in a complexity class that is considered feasible.

Besides, we consider a wider class of interaction: $|f(\vec{r_i},\vec{r_j})|\le ||\vec{r_i}-\vec{r_j}||^{-\alpha},\alpha>0$. We will still call this family of interaction $\alpha$-long-range interaction because the interaction is still dominated by $r^{-\alpha}$.

\subsection{Computational Complexity Classes for Optimization Problems}

The decision problems are usually in the following form: given the instance of the problem, the algorithm is required to answer ``yes'' or ``no''. For example, given a Hamiltonian of Ising model and energy $E$, returning whether this Ising model have a state with energy less than $E$ is an decision problem. Optimization problems, however, requires the algorithm to give an ``solution'' for the problem that is close enough to the ideal solution, for example, given a graph and a real number $\epsilon$, returning an independent subset $V'$ so that $|V'|\ge (1-\epsilon) |V|$ where $V$ is the maximal independent set of the graph is an optimization problem. The $\epsilon$-approximation requirement of the solution is used for definitions of some important complexity classes and a lot of practical problems have nice structures so that $\epsilon$-approximation can be defined~\cite{optimization_problems}, so do finding the ground state of Ising models, which is defined in the next section.

Some important complexity classes for optimization problems are defined as follows~\cite{optimization_classes}:

\begin{itemize}
    \item $\mathsf{APX}$ (approximable) is defined as all the NP optimization problems where there exists a polynomial-time approximation algorithm to give an $\epsilon$-approximation solution for constant epsilon.
    \item $\mathsf{PTAS}$ (Poly-Time Approximation Scheme) is defined as all of the NP optimization problems that have an approximation algorithm that returns an $\epsilon$-approximation for any given $\epsilon$ and the time complexity of the algorithm is polynomial to the problem length $n$. The exponent of the polynomial may still depends on $\epsilon$. For example, a problem that has an algorithm that gives $\epsilon$-approximation within $O(n^{\frac{1}{\epsilon!}})$ time is still considered in $\mathsf{PTAS}$, although good approximations are not feasible to compute.
    \item $\mathsf{FPTAS}$ (Fully Poly-Time Approximation Scheme) is defined as all of the NP optimization problems that have an approximation algorithm that returns an $\epsilon$-approximation for any given $\epsilon$ and the time complexity of the algorithm is polynomial to $n$ and $\frac{1}{\epsilon}$.
\end{itemize}

The relations between these complexity classes are still an open problem, but unless $\mathsf{P} = \mathsf{NP}$, we have $\mathsf{FPTAS} \subsetneq \mathsf{PTAS} \subsetneq \mathsf{APX}$~\cite{approximation_inclusion}.

\subsection{Definition of $\epsilon$-approximation}

For an Ising model with $n$ spins, we denote its ground state by $\sigma_0$. An state is an $\epsilon$-approximation if:

$$
E(\sigma)-E(\sigma_0) \le n\epsilon
$$

That is, each spin contributes $\epsilon$ error in energy on average.

The reason why we choose this definition is that it is consistent with the optimization version of the independent set problem of the graph that is discussed before: consider a planar graph with max degree 3 on which we want to find the maximum independent set. We can construct an Ising model as Lemma~\ref{thm:nn}. Suppose the cardinality of the maximum independent set is $n_0$, which is greater than $\frac{|V|}{4}$ because the graph have max degree 3. Then the corresponding ground state of the Ising model has energy $\frac{1}{2}|V| -4(n_0+1)\le E_0\le \frac{1}{2}|V| - 4n_0$. Consider a state with energy $E\le E_0+n\epsilon$, so its energy is $E\le \frac{1}{2}|V|-4n_0(1-\epsilon)$, which indicates an independent set is $(1-\epsilon)n_0$, which is an $\epsilon$-approximation of independent set problem.

\subsection{Ground state of tree-like Ising models}\label{sec:dp}

The tree decomposition groups vertices of a graph into bags that have a tree structure, which is often used to accelerate solving some problems on graph. The definition of the tree decomposition is: \cite{graph_tree_dp}

For a graph $G=(V,E)$, a tree decomposition of the graph is $(X,T)$, where each element $X_i\in X$ is a subset of $V$ and $T$ is a tree with nodes in $X$. They satisfies:

\begin{enumerate}
    \item $\bigcup_i X_i = V$, that is each node in $v$ is in at least one bag.
    \item If there exists an edge $\{u,v\}$, then there exist an $X_i$ such that both $u$ and $v$ are in $X_i$.
    \item If both $X_i$ and $X_j$ contains $v$, then every node on the unique path between $X_i$ and $X_j$ in $T$ contains $v$ as well.
\end{enumerate}

The width of the decomposition is defined as the size of the largest bag minus 1, and the treewidth of the graph is defined as the minimum width for all the tree decompositions. Treewidth is a measure of how close the graph is to a tree: A tree clearly has treewidth 1, and any graph that contains a cycle has treewidth at least 2. Notice that there exists no upper bound for planar graph as the treewidth of an $n\times n$ square grid is $n$. Some problems on a graph with a low treewidth can be solved efficiently by dynamic programming, and solving the ground state of an Ising model is one of them.

For a tree decomposition $(X,T)$, we define an \emph{assignment} $H$ as a set of maps $h_i:X_i\rightarrow \{-1,+1\}$. An assignment is \emph{consistent} if for any vertex $v\in V$, if $v\in X_i$ and $v\in X_j$, then $h_i(v)=h_j(v)$. An consistent assignment will give a vertex only one value no matter which bag it is in, so a consistent assignment corresponds a state.

We can find the ground state of an Ising model given its tree decomposition by the algorithm:~\cite{graph_tree_dp}

Picking a bag $X_i$, we enumerate all of its assignment $h_i$ and want to find the lowest energy assignment consistent with it. We can remove $X_i$ from $T$ and obtain an forest $\{T_j\}$, and denote by $X_j$ the (bag) vertices that are adjacent to $X_i$ in the graph and lies in $T_j$. For each $T_j$, a part of its vertices has been assigned, and we use this algorithm recursively to solve the ground state of $T_j$ given these vertices fixed, which gives the lowest energy assignment of $T$ given $h_i$ fixed. Finally, the ground state, is given by the lowest energy assignment along all $h_i$.

Suppose the width of the tree decomposition is $D$, then with dynamic programming technique, solving the ground state takes $O(2^D |X|)$ time.

\subsection{Approximation scheme}

Now, given a long-range interaction, we want to prune the interaction between some spins to get a low treewidth graph, while introducing as small errors to energy of states as possible.

For an RFIM, the interactions between the spins can be denoted by a graph $G=(V,E)$, with the interaction between all neighboring vertices being counted. We define an approximation graph $\tilde{G}=(V, \tilde{E})$ as a subgraph of $G$, and we can construct another Hamiltonian of the system on this approximation graph:

$$
H_{\tilde{G}} = -\sum_{\{i, j\}\in \tilde{E}} J_{ij} \sigma_i\sigma_j - \sum_{k\in V} h_k\sigma_k
$$

If an approximation graph satisfies: $\forall \sigma, |H(\sigma) - H_{\tilde{G}}(\sigma)|\le  \frac{1}{2} n\epsilon$, then we call this graph an \emph{$\epsilon$-approximation graph}.

The following theorem gives an sufficient condition for the possibility of efficiently solving an approximation:

\begin{lem}\label{thm:apx}
For an RFIM, if there exist an $\epsilon$-approximation graph that have a treewidth $D$, then given this minimum width tree decomposition, the algorithm described in Section~\ref{sec:dp} will give an $\epsilon$-approximation under $O(2^D |X|)$ time.
\end{lem}

Following the theorem above, we can obtain an sufficient condition for a problem in $\mathsf{PTAS}$ immediately:

\begin{thm}\label{thm:ptas}
For a problem (that solves approximation of RFIM), if for all instances, let $n$ be the size of the instance, then if for an arbitrarily small $\epsilon>0$, this instance has an $\epsilon$-approximation graph, and this graph have a tree decomposition $(X,T)$, with $|X|$ being a polynomial of $n$ and the upper-bound of width of the decomposition $D$ is a constant to $n$, then this problem is in $\mathsf{PTAS}$.
\end{thm}

Using this theorem, we can show that approximating the 1-dimensional $\alpha$-long-range interaction RFIM is in $\mathsf{PTAS}$ with $\alpha\ge 1$:

\begin{cor}\label{thm:ptas1d}
For a 1-dimensional $\alpha$-long-range interaction RFIM ($n$ spins) with every spin on the integer coordinate, it has an $\epsilon$-approximation graph $\tilde{G}$ that excludes edges between vertices with distance $>(\epsilon(\alpha-1))^\frac{1}{1-\alpha}=R$. There exists a tree decomposition of this approximation graph $(X,T)$ that $X_i = \{v_i,\cdots,v_{i+R}\}$ and $X_i,X_{i+1}$ are connected in $T$. The width of this decomposition is constant to $n$ and $|X|=O(n)$, hence this Ising model is in $\mathsf{PTAS}$.
\end{cor}

This theorem does not apply for $\alpha\le 1$ as in this scenario, the energy mainly consists of interactions between spins that are far away from each other.

There is an intuition that 2-dimensional square-grid long-range RFIM is not in $\mathsf{PTAS}$, due to the following observation: the edges of the square grid consists of interaction of strength $1$, which implies that an approximation graph excluding a lot of square grid edges is not likely to be an $\epsilon$-approximation graph with $\epsilon<1$. So, it is not likely that for a fixed $\epsilon<1$, there exists an $\epsilon$-approximation graph with fixed graph width, which implies that 2-dimensional square-grid long-range RFIM is probably not in $\mathsf{PTAS}$.

\begin{acknowledgments}
We thank Y.-K. Wu for discussions. This work was supported by the Frontier Science Center for Quantum Information of the Ministry of Education of China and the Tsinghua University Initiative Scientific Research Program.
\end{acknowledgments}

\appendix

\section{Appendix A: Proof for the theorems}

\begin{proof}[Proof for Lemma~\ref{thm:gap}]
There is:

$$
\begin{aligned}
&\min_{\sigma' \notin g^{-1}(G)} E(\sigma') - \max_{\sigma' \in g^{-1}(G)} E(\sigma')\\
\ge{}& a(\min_{\sigma \notin G} E(\sigma) - \delta - \max_{\sigma \in G} E(\sigma)-\delta)\\
={}& a(\Delta_G - 2\delta)
\end{aligned}
$$

\end{proof}

\begin{proof}[Proof for Lemma~\ref{thm:composite}]
There is:

$$
\begin{aligned}
 & |E(\sigma'') - a'aE(g(g'(\sigma'')))|\\
 \le{}& |E(\sigma'') - a'E(g'(\sigma''))| + a'|E(g'(\sigma'')) - a''E(g(g'(\sigma'')))|\\
 ={}& a'\delta' + a'a\delta\\
 ={}& a'a(\frac{\delta'}{a} + \delta)
\end{aligned}
$$

So the composite map $g\circ g'$ is a $(a'a,\frac{\delta'}{a} + \delta)$-map.
\end{proof}

\begin{proof}[Proof for Lemma~\ref{thm:mapgrid}]

For any two spins, let $a, b$ be the position of them on the original system, and let $a', b'$ be the position of the two corresponding constructed spins. There is ($d(x,y)=||x-y||$):

$$
\begin{aligned}
\frac{1}{\epsilon} d(a, b) -2 \le d(a',b') \le \frac{1}{\epsilon} d(a, b) + 2\\
\end{aligned}
$$

And there is, under $|x|<\frac{t}{2(\alpha+1)}<\frac{1}{2(\alpha+1)}$:

$$
|(1+x)^{-\alpha}-1| \le \alpha (1-x)^{-\alpha-1} |x|\le t
$$

The Hamiltonian of the system is:

$$
\begin{aligned}
H'(\sigma')=&-\sum_{a',b'\in S'} J_{a'b'} \sigma_{a'}\sigma_{b'} - \sum_{c'\in S'} h'_{c'} \sigma_{c'}\\
=&-\sum_{a', b'\in S'} -d^{-\alpha}(a',b') \sigma_{a'} \sigma_{b'} - \sum_{c'\in S'} h_{c} \epsilon^\alpha \sigma_{c'}\\
\end{aligned}
$$

So:

$$
\begin{aligned}
& \left| H'(\sigma')+(\sum_{a',b'\in S}-d^{-\alpha}(a,b)\sigma_{a'}\sigma_{b'}+\sum_{c'\in S'} h_{c} \sigma_{c'})\epsilon^\alpha\right| \\
=& \left|\sum_{a',b'\in S'} (d^{-\alpha}(a',b')-\epsilon^\alpha d^{-\alpha}(a,b) ) \sigma_{a'}\sigma_{b'}\right|\\
\le & \sum_{a', b'} \left|  d^{-\alpha}(a',b')-\epsilon^\alpha  d^{-\alpha}(a,b)  \right|\\
=& \sum_{a', b'} \epsilon^\alpha d^{-\alpha}(a, b) \left| (\frac{\epsilon d(a', b')}{d(a,b)})^{-\alpha} -1\right|\\
\le & \sum_{a', b'} \epsilon^\alpha d^{-\alpha}(a, b) \left| (1+\frac{\epsilon d(a', b')-d(a,b)}{d(a,b)})^{-\alpha} -1\right|\\
\end{aligned}
$$

Now if we set $\epsilon \le \frac{t d}{4(\alpha+1)}$ with $t\in(0,1]$, then for all $a', b'$, there is:

$$
\left|\frac{\epsilon d(a', b')-d(a,b)}{d(a,b)}\right|\le \frac{2\epsilon}{d(a,b)} \le \frac{t}{2(\alpha+1)}
$$

There is:

$$
\begin{aligned}
&|H'(\sigma') - \epsilon^\alpha H(g(\sigma'))|\\
=& \left| H'(\sigma')+(\sum_{a',b'\in S}-d^{-\alpha}(a,b)\sigma_{a'}\sigma_{b'}+\sum_{c'\in S'} h_{c} \sigma_{c'})\epsilon^\alpha\right| \\
\le & \sum_{a', b'} \epsilon^{\alpha}d^{-\alpha}(a,b) t\\
\le & \epsilon^{\alpha} \frac{n(n+1)}{2} t d^{-\alpha}
\end{aligned}
$$

By setting $t\le \frac{2\delta d^\alpha}{n(n+1)}$, which implies $\epsilon<\frac{d^{\alpha+1} \delta}{2(\alpha+1)n(n+1)}$ we can conclude that $g$ is a $(\varepsilon^\alpha, \delta)$ mapping.
\end{proof}

\begin{proof}[Proof for Lemma~\ref{thm:valid}]
The interaction of spins on the logical spin satisfies the condition that: the total disturbance of the other spins together with the applied field is smaller than $|h|+8n(R-\sqrt{2})^{-\alpha} < \frac{\Delta}{2}$, so $\{\sigma_+, \sigma_-\}$ remains a set of low-energy state.
\end{proof}

\begin{proof}[Proof for Lemma~\ref{thm:interaction}]

Let $L$ and $L'$ denote the sets that contain the spins in two logical spins respectively.

The Hamiltonian of the system writes:

$$
\begin{aligned}
H(\sigma) =& -\sum_{a,b\in L\cup L'} -J_{a,b}\sigma_a\sigma_b - \sum_{c} h_c\sigma_c\\
=&-\sum_{a,b \in L} J_{a,b}\sigma_a\sigma_b - \sum_{a,b \in L'} J_{a,b}\sigma_a\sigma_b\\
&- \sum_{c\in L} h_c\sigma_c - \sum_{c\in L'} h_c\sigma_c - \sum_{a\in L, b\in L'} J_{a,b}\sigma_a\sigma_b\\
=&2C -\tilde{h_1}\tilde{\sigma_1} - \tilde{h_2}\tilde{\sigma_2} - I_{12} \tilde{\sigma_{1}}\tilde{\sigma_2}\\
\end{aligned}
$$
Where $C$ is a function of $\alpha$.

Now calculate $I_{12}$, the Taylor expansion of $J_{a,b}$ to the 4th order term shows that:

$$
I_{12} = -\alpha^2(\alpha+2)^2 r^{-(\alpha+4)} (1+O(r^{-1}))
$$

So we can conclude that:

$$
\left|I_{12} +\alpha^2(\alpha+2)^2 r^{-(\alpha+4)}\right| \le f(\alpha) r^{-\alpha-5}
$$

\end{proof}

\begin{proof}[Proof for Lemma~\ref{thm:maplogical}]
Using Lemma~\ref{thm:interaction}, the Hamiltonian of the constructed system is, up to a constant:

$$
\begin{aligned}
H'(\Tilde{\sigma}) =& -\sum_{c} \tilde{h_c} \tilde{\sigma_c} -\sum_{a,b} I_{a,b} \tilde{\sigma_a}\tilde{\sigma_b}
\end{aligned}
$$

Let $\sigma = g(\Tilde{\sigma})$, there is:

$$
\begin{aligned}
&\left| H'(\Tilde{\sigma}) - \alpha^2(\alpha+2)^2 \epsilon^{\alpha+4} H(\sigma) \right|\\
={}&\left| \sum_c \tilde{h}_c\tilde{c} + \sum_{a,b} I_{a,b} \tilde{\sigma_a}\tilde{\sigma_b}\right. \\&- \left.\sum_c \alpha^2(\alpha+2)^2 \epsilon^{\alpha+4} h_c \tilde{\sigma_c} - \sum_{a,b} -\alpha^2(\alpha+2)^2 \epsilon^{\alpha+4} \tilde{\sigma_a}\tilde{\sigma_b} \right|\\
={}&\left|\sum_{a,b} (I_{a,b} + \alpha^2(\alpha+2)^2 \epsilon^{\alpha+4} )\tilde{\sigma_a}\tilde{\sigma_b} \right|\\
\le{}& \frac{n(n+1)}{2} f(\alpha) \epsilon^{\alpha+5}
\end{aligned}
$$

By choose $\epsilon < \delta \frac{2\alpha^2(\alpha+2)^2}{n(n+1)f(\alpha)}$, $g$ is a $(\alpha^2(\alpha+2)^2\epsilon^{\alpha+4}, \delta)$-mapping.
\end{proof}

\begin{proof}[Proof for Lemma~\ref{thm:upperbound}]
The Hamiltonian of the constructed Ising model is:

$$
H'(\sigma') = -\sum_{u,v\in V} -r_{uv}^{-\alpha}\sigma'_u\sigma'_v - \sum_{w\in V} -\sigma'_w
$$

So:

$$
\begin{aligned}
&\left| H'(\sigma') - H(g(\sigma')) \right|\\
=&\left| \sum_{u,v\in V} r_{uv}^{-\alpha} \sigma'_u \sigma'_v - \sum_{(u,v)\in E} \sigma'_u\sigma'_v\right|\\
=&\left| \sum_{(u,v) \notin E} r_{uv}^{-\alpha} \sigma'_u\sigma'_v \right|\\
\le& \frac{|V|(|V|+1)}{2} 2^{-\frac{\alpha}{2}}\\
\le & \frac{1}{4}
\end{aligned}
$$
\end{proof}

\begin{proof}[Proof for Lemma~\ref{thm:apx}]
Suppose the algorithm find $\sigma$ and let $\sigma_0$ be the ground state. There is:

$$
\begin{aligned}
&H(\sigma) - H(\sigma_0)\\
={}& (H(\sigma) -  H_{\tilde{G}}(\sigma)) + (H_{\tilde{G}}(\sigma) -  H_{\tilde{G}}(\sigma_0) )+  (H_{\tilde{G}}(\sigma_0) - H(\sigma_0))\\
\le{}& \frac{1}{2}n\epsilon + 0 + \frac{1}{2}n\epsilon\\
\le{}& n\epsilon
\end{aligned}
$$
\end{proof}


\bibliography{main}

\end{document}